\RequirePackage{amsmath}
\documentclass[runningheads]{llncs}

\usepackage{tcolorbox}
\usepackage{etoolbox}

\newcounter{observationcounter}

\newif\ifEditMode

\definecolor{revcolor}{RGB}{0,138,230} % calm blue for revisions

 \EditModetrue
\usepackage{preamble}
\usepackage{aliases}

\usepackage{pifont}% http://ctan.org/pkg/pifont
\newcommand{\cmark}{\green{\ding{52}}}%
\newcommand{\xmark}{\red{\ding{56}}}%

\usepackage{caption}
\usepackage{subcaption}
\usepackage{siunitx}
\usepackage[table]{xcolor}

\renewcommand{\figureautorefname}{Fig.}
\renewcommand{\tableautorefname}{Tab.}

\begin{document}

\title{The Express Lane to Spam and Centralization: \\ An Empirical Analysis of Arbitrum's Timeboost}
%\title{End of the Block is the New Top of Block. How Timeboost Makes the Rich Richer}
%\title{Timeboost is not enough to protect against SPAM: An in-depth analysis}

\titlerunning{An Empirical Analysis of Arbitrum's Timeboost}

\author{Johnnatan Messias\inst{1}
\and Christof Ferreira Torres\inst{2}}

\institute{{Max Planck Institute for Software Systems (MPI-SWS)}
\and {INESC-ID/Instituto Superior Técnico (IST), University of Lisbon}}

\authorrunning{J. Messias and C. Ferreira Torres} 

\maketitle
\begin{abstract}
  DeFi applications are vulnerable to MEV, where specialized actors profit by reordering or inserting transactions. To mitigate latency races and internalize MEV revenue, Arbitrum introduced \textit{Timeboost}, an auction-based transaction sequencing mechanism that grants short-term priority access to an \textit{express lane}. In this paper, we present the first large-scale empirical study of Timeboost, analyzing over 48.5 million express lane transactions and 494 thousand auctions between January 2025 and April 2026. Our results reveal five main findings. First, express lane control is highly centralized, with three entities winning 99.74\% of auctions. Second, while express lane access provides earlier inclusion, it is currently mainly used for CEX-DEX arbitrage, limiting the value of priority access for other types of MEV extraction. Third, approximately 30\% of timeboosted transactions revert, indicating that Timeboost does not effectively mitigate MEV inefficiencies such as spam. Fourth, secondary markets for reselling express lane rights struggle to sustain themselves due to poor execution reliability and unsustainable economics. Finally, auction competition declined as secondary markets introduced dedicated CEX-DEX execution lanes, enabling pay-per-transaction instead of pay-per-round billing and leading to steadily reduced revenue for the Arbitrum DAO. Taken together, these findings show that Timeboost fails to deliver on its stated goals of fairness, decentralization, and internalizing long term revenue for the DAO. Instead, it currently allows for collusion and narrows adoption, highlighting the limitations of auction-based ordering as a mechanism for fair transaction sequencing in rollups.

\end{abstract}
%
%
%

%------------------------------------------------------------------------------

\section{Introduction}

\gls{DeFi} has introduced new paradigms for trading~\cite{Chemaya@FC24,Qin@SP22}, lending~\cite{bartoletti2021sok,Messias@FC23,Qin@IMC21}, and settlement~\cite{daian2020flash,Torres@USENIX21,Messias@IMC21,Qin@FC21}. However, transaction transparency and ordering enable \gls{MEV}, allowing actors to profit by reordering, inserting, or censoring transactions~\cite{daian2020flash,Torres@USENIX21,Weintraub@IMC22}. \gls{MEV} redistributes value from users to specialized searchers and threatens fairness, decentralization, and network stability~\cite{Torres@USENIX21,Qin@IMC21,Weintraub@IMC22}. Proposed mitigations range from cryptographic fair-ordering protocols such as Wendy~\cite{Kursawe2020Wendy}, Themis~\cite{Kelkar2023Themis}, and Helix~\cite{Yakira2021Helix} to \gls{PBS} and \gls{MEV} auctions~\cite{Heimbach@IMC23}.

At the \gls{L2} level, \gls{MEV} dynamics remain poorly understood~\cite{Torres@CCS24,gogol2024crossrollupmevnonatomicarbitrage}. Rollups such as Arbitrum~\cite{Kalodner-Usenix} process high transaction volumes through centralized sequencers, creating opportunities for latency-sensitive strategies and raising concerns about fairness, efficiency, and concentration. While centralized sequencing improves performance, it also concentrates control over transaction ordering and \gls{MEV} extraction.
To address these challenges, Arbitrum introduced \emph{Timeboost}~\cite{arbitrum2025timeboost}, an auction-based ordering mechanism that extends \gls{FCFS} with a market for sequencing priority. Timeboost divides time into one-minute rounds and allocates an express lane through a sealed-bid second-price auction. Winners obtain exclusive low-latency access, while regular transactions incur a deterministic 200ms delay. By replacing latency competition with auction-based allocation, Timeboost aims to reduce latency races and spam~\cite{solmaz2025optimistic}, internalize \gls{MEV} revenue for the DAO, and improve fairness. Winners can also resell access, creating potential secondary markets for ordering rights~\cite{Arbitrum-Risk-Analysis}.

Timeboost is one of the first large-scale production deployments of an auction market for transaction-ordering rights. Unlike proposals evaluated only theoretically or through simulations, it has processed hundreds of thousands of auctions and tens of millions of prioritized transactions, offering a unique opportunity to assess auction-based sequencing in practice.
However, auctions may trade latency races for new forms of competition favoring well-capitalized participants, raising entry barriers, and enabling strategic behavior. Whether auction-based sequencing improves market outcomes therefore remains an open question.

In this paper, we present the first large-scale empirical analysis of Timeboost, using Arbitrum transaction and auction data from January 1, 2025 to April 13, 2026, spanning \num{161244600} blocks and \num{1318618176} transactions, including Timeboost's launch on April 17, 2025.
This is the largest empirical study of an auction-based transaction-ordering mechanism to date.
We investigate four main questions: (i) Who captures express lane access, and does competition persist? (ii) How is express lane access used, and does it reduce failed execution? (iii) Which \gls{MEV} strategies benefit from prioritized ordering, and does access provide measurable economic gains? (iv) Does Timeboost generate sustainable revenue and support viable secondary markets?
We answer these questions by analyzing auction outcomes, bidding behavior, transaction ordering, reverted transactions, \gls{MEV} extraction, and secondary-market providers such as Kairos~\cite{Kairos}.

\noindent
\textbf{Contributions:} Our main contributions are summarized as follows:
\begin{itemize}
\item \textbf{Auction Concentration.} Despite rotating auctions, express lane access is highly concentrated. Three entities win over 99.74\% of rounds, showing that auction-based allocation does not ensure broad participation.

\item \textbf{Low Execution Reliability.} Around 30\% of timeboosted transactions revert. Priority access neither guarantees execution nor eliminates spam.

\item \textbf{Limited MEV Benefits.} 
Timeboost primarily serves CEX-DEX arbitrage, questioning the value of timeboost's early inclusion for atomic arbitrage.

\item \textbf{Secondary Market Failure.} 
Secondary market providers face high revert rates and low returns, limiting sustainability and efficiency sub-auctioning.

\item \textbf{Unsustainable Revenue.} 
While Timeboost generated overall \num{2442.84} ETH
for Arbitrum's DAO, declining competition reduced clearing prices and revenue as express lane access became concentrated and major players (i.e., Wintermute and Selini Capital) shifted to secondary markets such as Kairos.

\end{itemize}

\section{Background} \label{sec:background}

In this section, we provide preliminary background on Timeboost and Kairos.

\begin{figure}[tbh]
 \centering
 \includegraphics[width=\onecolgrid]{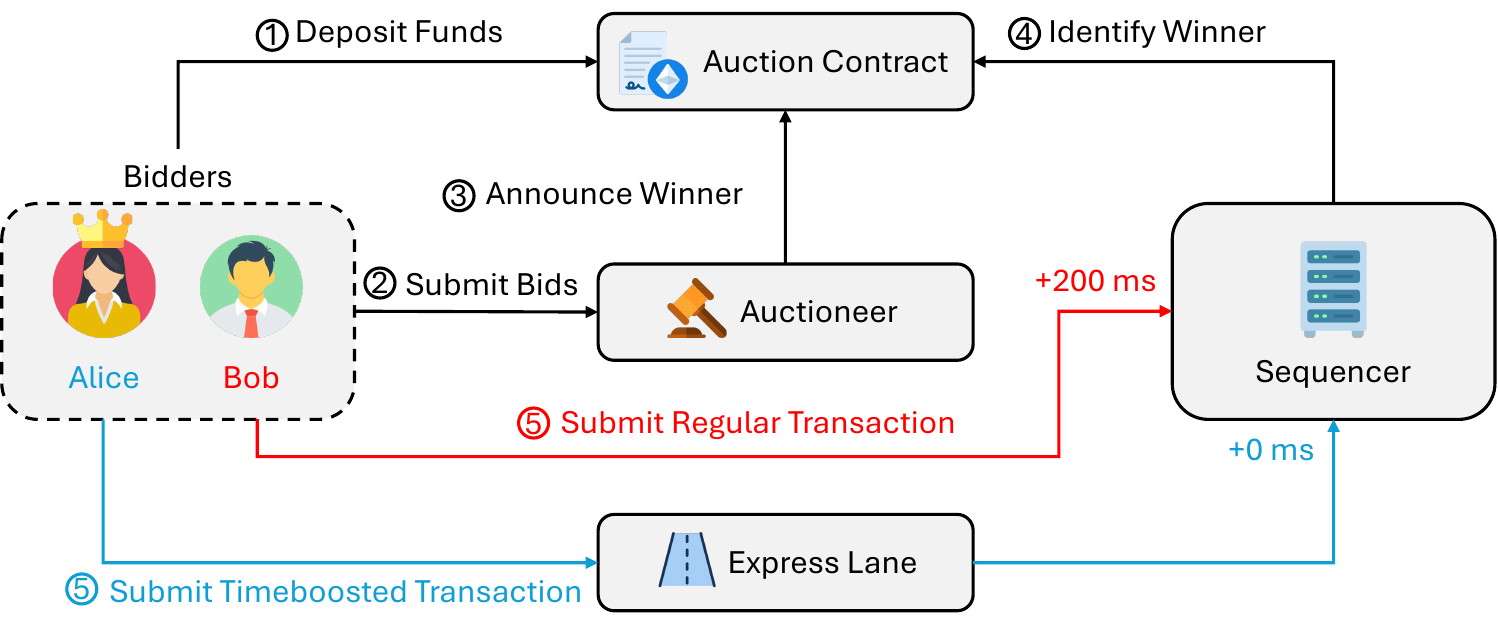}
 \caption{
 Overview of Timeboost’s architecture. Bidders deposit funds before submitting sealed bids to the auctioneer. After the auction, the auctioneer reports the winner to the auction contract, which the sequencer uses to grant express lane access. 
 Express lane transactions bypass the 200ms queuing delay (blue), while other transactions remain valid but are deterministically delayed (red).
  The winner (blue) may submit priority transactions, while other bidders (red) must submit regular transactions.
  Overview of Timeboost’s architecture. Bidders must first deposit funds before participating. They can then submit sealed bids to the auctioneer. Once the auction concludes, the auctioneer reports the winner to the auction smart contract. The sequencer verifies the winner via the contract and grants them access to the express lane. The winner (blue) is then able to submit priority transactions to the sequencer, while the loser (red) must continue submitting regular transactions.
 }
 \label{fig:timeboost-arch}
\end{figure}

\medskip
\noindent
\textbf{Timeboost.}
Timeboost divides time into 60-second rounds, each consisting of a \emph{pre-round} and \emph{execution} phase. During the pre-round phase, bidders deposit the minimum reserve price on-chain and submit sealed bids to the off-chain auctioneer for 45 seconds. The auctioneer selects the highest bidder and determines the clearing price as the second-highest bid, typically 15 seconds before the round starts. The outcome is recorded on-chain, designating the winner as the \emph{express lane controller}. During execution, the controller is the sole actor authorized to submit prioritized transactions through a dedicated RPC endpoint, while regular transactions use conventional endpoints and incur an artificial 200ms delay. Regardless of endpoint, transactions are ordered using FCFS.
Timeboost comprises five components: \emph{bidders}, an off-chain \emph{auctioneer}, an on-chain \emph{auction contract}, the \emph{express lane}, and the \emph{sequencer}. 
Fig.~\ref{fig:timeboost-arch} illustrates their interaction.
Bidders submit EIP-712 signed bids specifying the target round, bid amount, and controller address. The auctioneer ranks bids off-chain and submits the winner and clearing price to the auction contract, which records the controller and emits \texttt{SetExpressLaneController}. The sequencer monitors this event and authenticates express lane submissions. Specifically, transactions sent via \texttt{timeboost\_sendExpressLaneTransaction} must include the round number, per-round sequence number (\textit{nonce}), chain ID, and controller signature, together with the \gls{RLP}-encoded transaction payload. The sequencer verifies this metadata before inclusion.

\medskip
\noindent
\textbf{Kairos.} Gattaca's Kairos enables searchers to access the express lane directly without participating in Timeboost’s 60-second auctions. Kairos bids for round-level express lane access using a model that predicts the expected MEV value of each upcoming round, and shares this access with searchers through three modes: \emph{Standard Kairos}, which batches and simulates transactions in sub-auctions running roughly every 100 ms before ordering them by their offered payment; \emph{Optimistic Mode}, which bypasses simulation by accepting a declared payment upfront through an API header, reducing latency while retaining the same sub-auction mechanism; and \emph{Dedicated CEX/DEX Lanes}, which provide parallel pipelines with dedicated compute and network resources for latency-sensitive CEX-DEX arbitrage. Transactions submitted through the first two modes must include an explicit on-chain ETH payment to a dedicated Kairos address, whereas CEX-DEX lanes use a separate off-chain payment mechanism. For either mode, Kairos forwards eligible transactions through the express lane, allowing them to bypass Timeboost’s 200 ms delay and obtain prioritized sequencing, while charging searchers on a per-transaction basis rather than for the entire round.

\section{Data Collection} \label{sec:dataset}

To evaluate the impact of Timeboost in practice, we collect and curate multiple datasets from Arbitrum and Dune~\cite{Dune}. 
Table~\ref{tab:dataset} provides an overview of the datasets we collected, including their sources and number of items.
Our Timeboost data spans from January 1, 2025, (i.e., block \num{290688000}) to April 13, 2026, (i.e., block \num{451932599}), including  the introduction of Timeboost on April 17, 2025, (i.e., block \num{327331207}). Our dataset covers \num{161244600} blocks and \num{1318618176} transactions in total. As of the launch of Timeboost, we observe \num{1072698770} transactions included in \num{124601393} blocks in which \num{48525587} (4.52\%) are timeboosted. 
In addition, we also collect MEV related data such as cyclic/atomic arbitrages (i.e., DEX-DEX), non-atomic arbitrages (i.e., CEX-DEX), and liquidations.
A detailed description of the collection process for each dataset is provided in Appendix~\ref{sec:data-collection}\footnote{
We will release our datasets and scripts to support reproducibility and future work.
}.

\begin{table*}[t]
\centering
\caption{Overview of our collected datasets and items grouped by category.
}
\label{tab:dataset}
\begin{tabularx}{\textwidth}{@{}lXr@{}}
\toprule
\textbf{Dataset} & \textbf{Source} & \textbf{Items} 
\\ \midrule
\textbf{Arbitrum} & \\
\hspace{0.3cm} Blocks & Arbitrum Archive Node & \num{161244600} 
\\
\hspace{0.3cm} Transactions & Arbitrum Archive Node & \num{1318618176} 
\\
\textbf{Timeboost} \\
\hspace{0.3cm} Timeboosted Transactions & Arbitrum Archive Node & \num{48525587} 
\\
\hspace{0.3cm} Reverted Timeboosted Transactions & Arbitrum Archive Node & \num{14796548} 
\\
\hspace{0.3cm} Timeboost Auctions & Arbitrum API & \num{494608} 
\\
\hspace{0.3cm} Auction Winners & Arbitrum API & \num{32} 
\\
\hspace{0.3cm} Auction Bidders & Arbitrum API & \num{40} 
\\
\textbf{MEV Extraction} \\
\hspace{0.3cm} Atomic/Cyclic Arbitrages (DEX-DEX) & Arbitrum Archive Node & \num{7779084} \\
\hspace{0.3cm} Non-Atomic Arbitrages (CEX-DEX) & Arbitrum Archive Node & \num{86877427} \\
\hspace{0.3cm} Liquidations & Arbitrum Archive Node & \num{20299} 
\\
\textbf{Kairos Revenue} \\
\hspace{0.3cm} Kairos Payment Transactions & Dune Query & \num{349076} 
\\
\bottomrule
\end{tabularx}
\end{table*}

\parab{Limitations} 
Our measurements should be interpreted as a lower bound, as we do not cover all protocols available on Arbitrum. We currenlty cover Aave, Balancer V1-V2, Curve, Compound, Uniswap V2-V4, PancakeSwap V3.
Moreover, similar to prior work~\cite{oz2026jit}, our non-atomic arbitrage detection is restricted to single-swap arbitrages involving only WETH, WBTC, USDC, and USDT.

\section{General Analysis of Timeboost} \label{sec:analysis}

This section examines express lane control, adoption, Timeboost usage, auction dynamics, reverted transactions, and revenue.

\subsection{Express Lane Controller Share}

\begin{figure*}[tbh]
  \centering
\includegraphics[width=\textwidth]{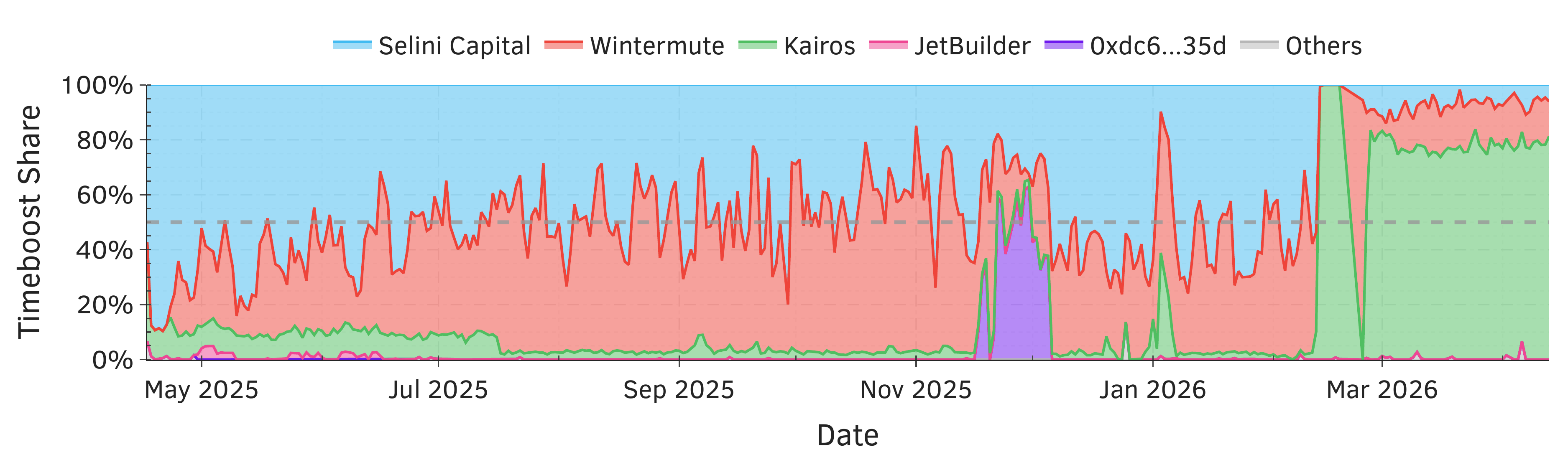}
  \caption{Controller share from April 17, 2025, to April 13, 2026. Selini Capital, Wintermute, and Kairos collectively controlled 99.74\% of winning rounds.
  }
  \label{fig:timeboost-share}
\end{figure*}

As shown in Fig.~\ref{fig:timeboost-share}, control is highly concentrated. Among 32 auction winners, three entities, Selini Capital, Wintermute, and Kairos, accounted for 99.74\% of rounds, with Selini and Wintermute alone accounting for 77.41\%. This concentration suggests that Timeboost disproportionately favors sophisticated actors with optimized bidding strategies or greater capital. From February 13, 2026 onward, Kairos dominated with an average share of 90.96\%. Thus, although Timeboost rotates express lane control across rounds, \stress{access is effectively centralized}, raising concerns about fairness and accessibility for smaller or new participants.

\subsection{Timeboost Adoption}

\begin{figure*}[tbh]
  \centering
\includegraphics[width=\textwidth]{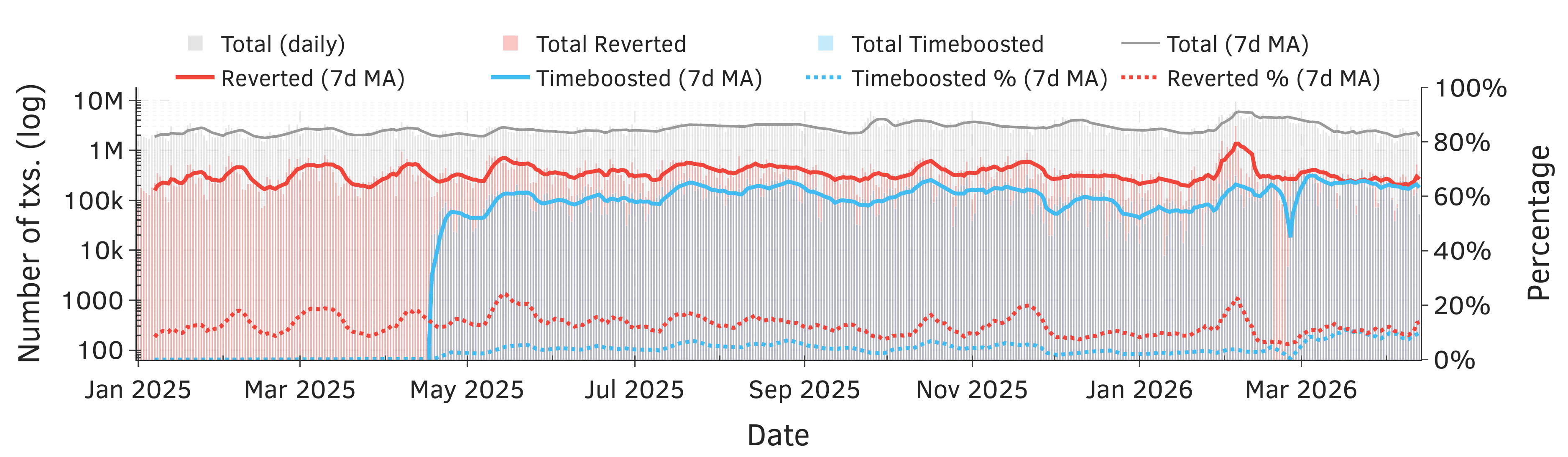}
  \caption{
  Timeboost usage remains volatile, accounting for 0--16.05\% of all transactions, indicating strategy-driven adoption, while revert ratio remains identical.
  }
  \label{fig:timeboost-transactions}
\end{figure*}

\begin{figure*}[tbh]
  \centering
\includegraphics[width=\textwidth]{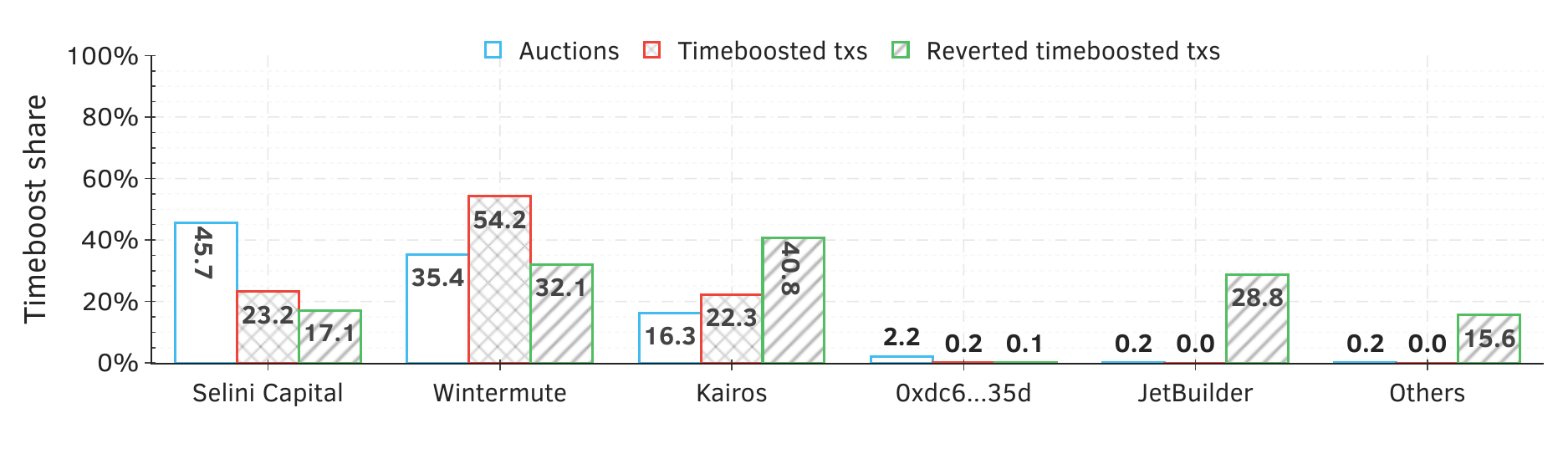}
  \caption{
  Timeboost activity per controller: auction wins (solid), timeboosted transactions (cross-hatched), and reverted timeboosted transactions (dashed). Note that reverted shares are controller-specific and therefore do not sum to 100\%.
  }
  \label{fig:timeboost-txs-auction-share-comparison}
\end{figure*}

Fig.~\ref{fig:timeboost-transactions} shows that timeboosted transactions constitute a small and volatile share of Arbitrum activity, suggesting episodic, strategy-driven usage such as \gls{MEV}. Their daily share ranges from 0 to 16.05\% without a persistent trend. Controller behavior is similarly concentrated (see Fig.~\ref{fig:timeboost-txs-auction-share-comparison}). Although Selini Capital wins 45.7\% of auctions, Wintermute submits 54.2\% of timeboosted transactions versus Selini's 23.2\%, indicating greater express lane utilization despite fewer wins.
Overall, timeboosted transactions target 165 unique `to' addresses, 153 of which are contracts. 
\tableautorefname{}~\ref{tab:most-freq-timeboost-receivers} shows the top ten recipients of timeboosted transactions.
The \stress{top three contracts account for 88.91\% of all timeboosted transactions, primarily driven by CEX-DEX arbitrage.} Several smaller recipients, not listed in \tableautorefname{}~\ref{tab:most-freq-timeboost-receivers}, exhibit revert rates exceeding 70--90\%, indicating that while most Timeboost traffic flows to a small set of contracts, many other timeboosted contracts still suffer from poor execution reliability.
We also find that nine of the top ten contracts are associated with Selini or Wintermute due to their exclusive or substantial controller share. Transactions initially sent exclusively to Selini or Wintermute shifted toward Kairos around mid-February 2026, suggesting that these entities began using Kairos' controller to access the express lane, likely following Kairos’ introduction of dedicated CEX-DEX lanes.

\begin{table}[th]
    \centering
        \caption{
        Top 10 contracts receiving timeboosted transactions sorted by frequency.
        Rows highlighted in gray highlight CEX-DEX arbitrage activity.
        }
    \begin{adjustbox}{max width=\columnwidth}
    \begin{tabular}{l r r c r r c }
    \toprule
    \textbf{Contract} & \multicolumn{2}{c}{\textbf{Timeboosted Txs}} & \textbf{MEV Type} & \textbf{Reverted} & \multicolumn{1}{c}{\textbf{Activity Period}} & \textbf{Controller Share*} \\
    \midrule
    \cellcolor{gray!20}\href{https://arbiscan.io/address/0x27920e8039d2b6e93e36f5d5f53b998e2e631a70}{\texttt{0x279$\cdots$a70}} & \cellcolor{gray!20}\num{26148218} & \cellcolor{gray!20}(53.89\%) & \cellcolor{gray!20}CEX-DEX Arb. & \cellcolor{gray!20}36.11\%  & \cellcolor{gray!20} 2025-08-04  2026-04-13 & \cellcolor{gray!20} \textbf{\textit{W}} (74\%), \textbf{\textit{K}} (26\%) \\

    \cellcolor{gray!20}\href{https://arbiscan.io/address/0xee2e7bbb67676292af2e31dffd1fea2276d6c7ba}{\texttt{0xee2$\cdots$7ba}} & \cellcolor{gray!20}\num{10017621} & \cellcolor{gray!20}(20.64\%) & \cellcolor{gray!20}CEX-DEX Arb. & \cellcolor{gray!20}13.56\%  & \cellcolor{gray!20}2025-04-17 2026-04-13 & \cellcolor{gray!20} \textbf{\textit{S}} (88\%), \textbf{\textit{K}} (12\%) \\
    
    \cellcolor{gray!20}\href{https://arbiscan.io/address/0xcb43d843f6cadf4f4844f3f57032468aadd9b95c}{\texttt{0xcb4$\cdots$95c}} & \cellcolor{gray!20}\num{6980181} & \cellcolor{gray!20}(14.38\%) & \cellcolor{gray!20}CEX-DEX Arb. & \cellcolor{gray!20}27.26\%  & \cellcolor{gray!20} 2025-04-23 2025-08-04 & \cellcolor{gray!20} \textbf{\textit{W}} (100\%) \\

   \cellcolor{gray!20}\href{https://arbiscan.io/address/0x96daa0b8a5499ea9323421ed0cda06b345caab73}{\texttt{0x96d$\cdots$b73}} & \cellcolor{gray!20}\num{1460856} & \cellcolor{gray!20}(3.01\%) & \cellcolor{gray!20}CEX-DEX Arb. & \cellcolor{gray!20}40.42\%  & \cellcolor{gray!20}2026-02-13 2026-04-13  & \cellcolor{gray!20} \textbf{\textit{K}} (100\%) \\
    
    \href{https://arbiscan.io/address/0x00000000112b9395150ca682820869b2260b66a4}{\texttt{0x000$\cdots$6a4}} & \num{1156241} & (2.38\%) & DEX-DEX Arb. & 38.10\%  &  2025-07-15 2025-11-22 & \textbf{\textit{S}} (99\%), \textbf{\textit{K}} (1\%)   \\

    \cellcolor{gray!20}\href{https://arbiscan.io/address/0x78fe61cf0124fa08c4b79d5ac47ff8f152569535}{\texttt{0x78f$\cdots$535}} & \cellcolor{gray!20}\num{538817} & \cellcolor{gray!20}(1.11\%) & \cellcolor{gray!20}CEX-DEX Arb. & \cellcolor{gray!20}9.22\%  & \cellcolor{gray!20}2025-11-05 2026-03-25 & \cellcolor{gray!20}  \textbf{\textit{S}} (56\%), \textbf{\textit{K}} (44\%)  \\

    \cellcolor{gray!20}\href{https://arbiscan.io/address/0x79d252a8dfae2ca79ba98dc963aa025109e136f6}{\texttt{0x79d$\cdots$6f6}} & \cellcolor{gray!20}\num{252742} & \cellcolor{gray!20}(0.52\%) & \cellcolor{gray!20}CEX-DEX Arb. & \cellcolor{gray!20}18.31\%  & \cellcolor{gray!20}2025-10-03 2026-03-07 & \cellcolor{gray!20} \textbf{\textit{S}} (88\%), \textbf{\textit{K}} (12\%)   \\

    \href{https://arbiscan.io/address/0x00000000280b79b3aa82ec2a795582d1be1fc159}{\texttt{0x000$\cdots$159}} & \num{236552} & (0.49\%) & DEX-DEX Arb. & 40.45\%  & 2025-10-29 2026-02-17  &  \textbf{\textit{S}} (97\%), \textbf{\textit{K}} (3\%)  \\

    \href{https://arbiscan.io/address/0x8a1ba3d5b7864621a6214627a85a3f252b2e6180}{\texttt{0x8a1$\cdots$180}} & \num{223050} & (0.46\%) & DEX-DEX Arb. & 33.62\%  & 2026-03-06 2026-04-13 &  \textbf{\textit{S}} (6\%), \textbf{\textit{K}} (94\%)  \\

    \href{https://arbiscan.io/address/0xaac629672eff88119ed0c8a55f2e50658f7cf9d1}{\texttt{0xaac$\cdots$9d1}} & \num{147386} & (0.30\%) & DEX-DEX Arb. &  91.80\% &   2025-06-04 2025-09-24 &  \textbf{\textit{S}} (100\%) \\

\bottomrule
\multicolumn{7}{l}{*\textbf{\textit{W}}: Wintermute, \textbf{\textit{S}}: Selini Capital, \textbf{\textit{K}}: Kairos}
    \end{tabular}
    \end{adjustbox}
    \label{tab:most-freq-timeboost-receivers}
\end{table}

\subsection{Reverted Transactions} 

\begin{figure*}[tbh]
  \centering
\includegraphics[width=\textwidth]{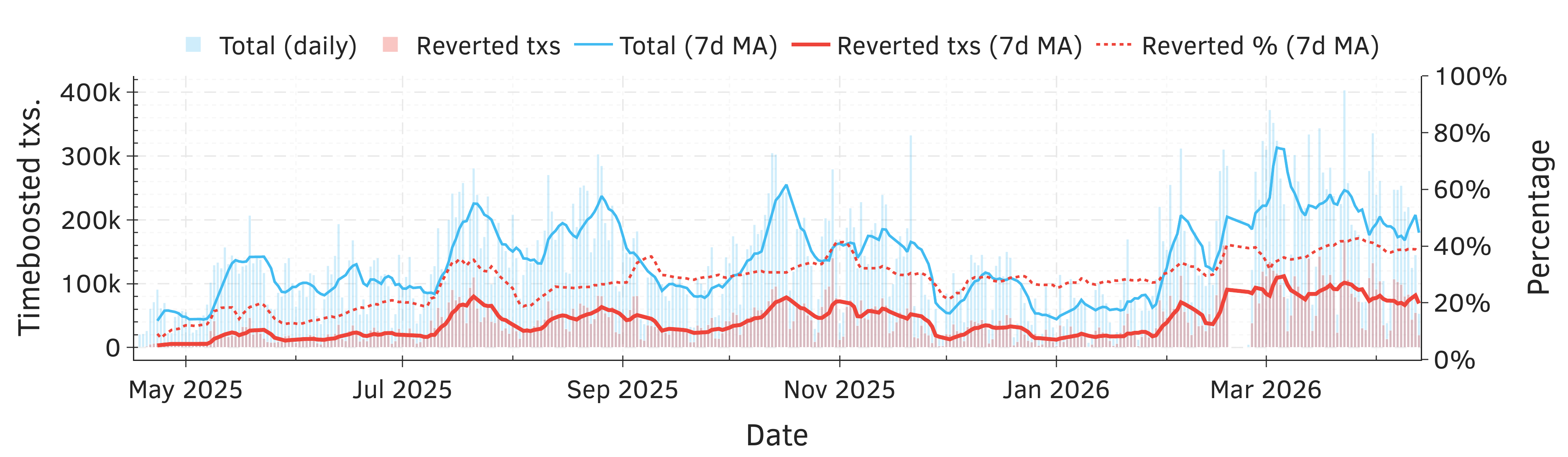}
  \caption{
  Daily timeboosted transactions and reverted timeboosted transactions with 7-day moving averages. 30.49\% of timeboosted transactions revert.
  }
  \label{fig:timeboost-reverted-txs}
\end{figure*}

\begin{figure*}[tbh]
  \centering
\includegraphics[width=\textwidth]{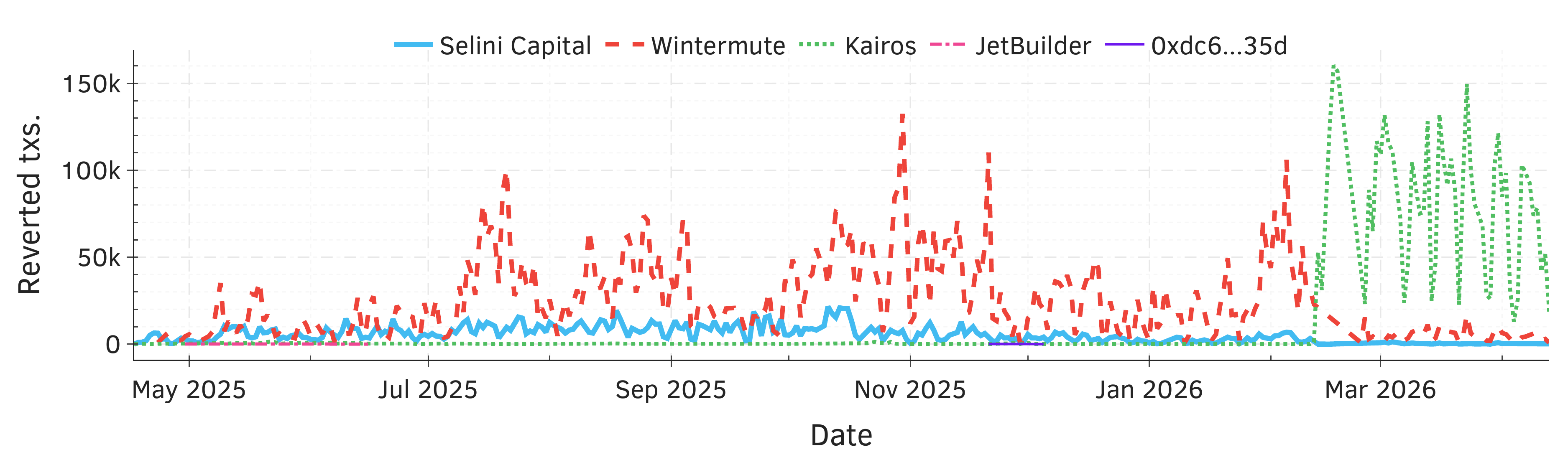}
  \caption{
  Reverted timeboosted transactions over time. Wintermute initially dominated, but Kairos took over from February 2026 by securing more auctions.
  }
  \label{fig:timeboost-reverted-txs-overtime}
\end{figure*}

Before Timeboost, 13.83\% of Arbitrum transactions reverted, compared to 12.61\% after its launch (11.23\% regular and 1.38\% timeboosted). While this suggests Timeboost may have reduced spam~\cite{solmaz2025optimistic}, \num{14796548} of \num{48525587} timeboosted transactions (30.49\%) still reverted.
Fig.~\ref{fig:timeboost-reverted-txs} shows daily reverts, 7-day moving averages, and their share. 
Revert rates also vary substantially (see Fig.~\ref{fig:timeboost-txs-auction-share-comparison}). Kairos has the highest rate at 40.8\%, followed by Wintermute (32.1\%) and JetBuilder (28.8\%), while smaller controllers average 15.6\%. These results show that \stress{express lane access does not guarantee successful execution}. Even with a 200ms advantage, strategy and execution quality remain decisive.  
A notable trend emerges after February 13, 2026, when Kairos became the dominant express lane controller (see Fig.~\ref{fig:timeboost-share}). During this period, both the number and share of reverted transactions increased substantially (see Figs.~\ref{fig:timeboost-reverted-txs} and \ref{fig:timeboost-reverted-txs-overtime}). As Kairos resells express lane access, intermediated access may introduce additional execution risk. Unlike direct winners, Kairos needs to coordinate transactions from multiple parties, potentially targeting the same opportunities.
Kairos’ documentation~\cite{Kairos-Docs} claims its bundling API provides Ethereum-style atomicity, excluding bundles if any transaction reverts. However, nearly half of Kairos transactions revert, contradicting this guarantee.
We replayed historical timeboosted transactions using Foundry~\cite{Foundry}. In block \num{363638889}, Kairos submitted six transactions: the first two succeeded, while the remaining four reverted. The successful transactions paid Kairos, whereas the reverted ones did not, despite being timeboosted (see Table~\ref{tab:kairos-reverted-txs}). When replayed independently at the top of the block, all six succeeded and paid Kairos. This suggests that Kairos processes bundles independently and timeboosts them without accounting for interdependencies, causing some to revert once sequenced by Arbitrum and contributing to the high failure rate.
Our results suggest that Timeboost does not necessarily reduce spam. If reverted transactions proxy unsuccessful MEV attempts, nearly one-third of express lane submissions waste blockspace. Rather than eliminating spam, Timeboost shifts it from the regular queue to the express lane, while, contrary to Kairos’ documentation, reverted transactions are being timeboosted for free.

\begin{table}[b]
    \centering
        \caption{Sample of reverted transactions submitted through Kairos.
        }
    \begin{adjustbox}{max width=\columnwidth}
    \begin{tabular}{c r r r c c c}
    \toprule
    \textbf{Transaction} & \textbf{Gas Limit} & \textbf{Gas Price} & \textbf{Index} & \textbf{Success} & \textbf{From} & \textbf{To} \\
    \midrule

\href{https://arbiscan.io/tx/0xa0db2445c926f628419c86996bea23b84be9bec62d81a8a4facdd374c1959482}{\texttt{0xa0d$\cdots$482}}
&
\num{3000000}
&
\num{11250000}
&		
2
&
\cmark
&	
\href{https://arbiscan.io/address/0x3c56b9a6dec1dbd034edd0b6846df8b0fef3b59c}{\texttt{0x3c5$\cdots$59c}}
& 
\href{https://arbiscan.io/address/0x381ede0cf36b56641209687ce28a84eaf3ac497e}{\texttt{0x381$\cdots$97e}}
\\

%$\cdots$ & $\cdots$ & $\cdots$ & $\cdots$ & $\cdots$ & $\cdots$ & $\cdots$ \\
    
\href{https://arbiscan.io/tx/0x82274d7772a4f57654ffa9e3173de432df8e6fef3cb3c6918045d81c21f6f012}{\texttt{0x822$\cdots$012}}
&
\num{3000000}
&
\num{11250000}
&	
7
&
\cmark
&
\href{https://arbiscan.io/address/0x7c5f8b294a68f3f381eb6d786bb1c67db96ef93e}{\texttt{0x7c5$\cdots$93e}}
&
\href{https://arbiscan.io/address/0x381ede0cf36b56641209687ce28a84eaf3ac497e}{\texttt{0x381$\cdots$97e}}
\\
	
\href{https://arbiscan.io/tx/0x745bb6a4bf56d48ea7503e3e81e57c90b28fdbac8149e5a8567f2ce0e9c29bd4}{\texttt{0x745$\cdots$bd4}}
&
\num{3000000}
&
\num{11250000}
&	
8
&	
\xmark
&	
\href{https://arbiscan.io/address/0x121eb2e23ebc09d1345101e5281e402c2c8001a0}{\texttt{0x121$\cdots$1a0}}
&
\href{https://arbiscan.io/address/0x381ede0cf36b56641209687ce28a84eaf3ac497e}{\texttt{0x381$\cdots$97e}}
\\

\href{https://arbiscan.io/tx/0xb61bcbb41e58eb9dd8f62c33de6540a61fa841d538fd382522e03b1ac93fb8f0}{\texttt{0xb61$\cdots$8f0}}
&
\num{3000000}
&
\num{11250000}
&
9
&
\xmark
&
\href{https://arbiscan.io/address/0x9c2a4f7c4cac453556a9418373a95e3c3ce0ac1d}{\texttt{0x9c2$\cdots$c1d}}
&
\href{https://arbiscan.io/address/0x381ede0cf36b56641209687ce28a84eaf3ac497e}{\texttt{0x381$\cdots$97e}}
\\

\href{https://arbiscan.io/tx/0xb404d8f1572087961c96317d071211b1e8ed00451010073f2d6648a4b464fbfc}{\texttt{0xb40$\cdots$bfc}}
&
\num{3000000}
&
\num{11250000}
&
24
&
\xmark
&
\href{https://arbiscan.io/address/0xa935547aebc5271e30bc35407a4d9e3c8879f618}{\texttt{0xa93$\cdots$618}}
&
\href{https://arbiscan.io/address/0x381ede0cf36b56641209687ce28a84eaf3ac497e}{\texttt{0x381$\cdots$97e}}
\\

\href{https://arbiscan.io/tx/0xb7f33d5d2533a2d3fb3858fa7be01ef17adc1130c6389f3404d239be66d678b5}{\texttt{0xb7f$\cdots$8b5}}
&
\num{3000000}
&
\num{11250000}
&
36
&
\xmark
&
\href{https://arbiscan.io/address/0x390aaba0cdd25583daefdb8a9204859c133204b3}{\texttt{0x390$\cdots$4b3}}
&
\href{https://arbiscan.io/address/0x381ede0cf36b56641209687ce28a84eaf3ac497e}{\texttt{0x381$\cdots$97e}}
\\

\bottomrule
    \end{tabular}
    \end{adjustbox}
    \label{tab:kairos-reverted-txs}
\end{table}

\subsection{Auction Activity}

\begin{figure*}[tbh]
    \centering
    \begin{subfigure}{\twocolgrid}
        \centering
        \includegraphics[width=\twocolgrid]{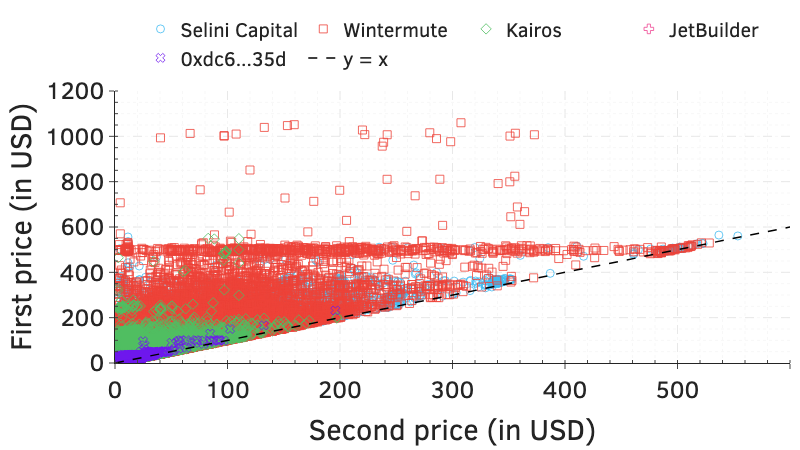}
        \caption{Winning bids vs. clearing prices.
        }
        \label{fig:timeboost-first-vs-second-price}
    \end{subfigure}
    \begin{subfigure}{\twocolgrid}
        \centering
        \includegraphics[width=\twocolgrid]{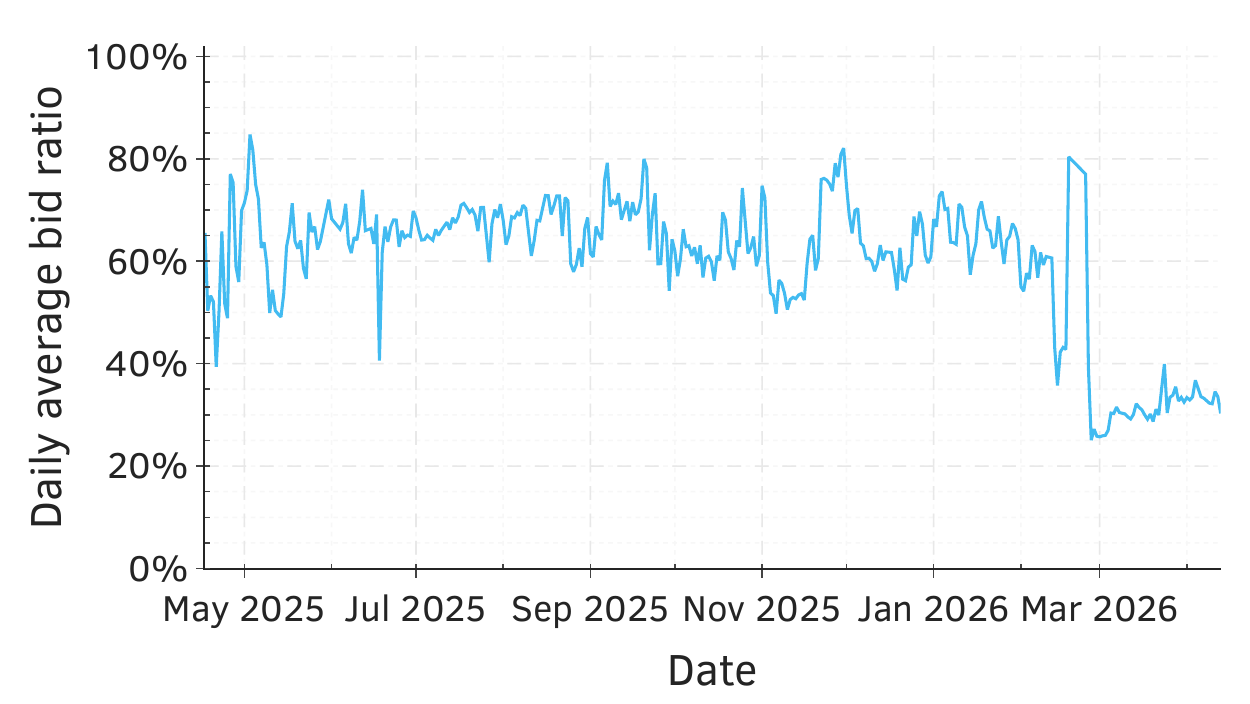}
        \caption{Daily average bid ratio.}
        \label{fig:timeboost-bid-ratio}
    \end{subfigure}\hfill
    \caption{
    Bidding behavior in Timeboost auctions. Winning bids consistently exceed clearing prices, with large entities systematically overbidding. Bid ratios stabilize as Selini Capital and Wintermute emerge as dominant controllers.
    }
    \label{fig:timeboost-behaviour}
\end{figure*}

Timeboost uses a sealed-bid second-price auction, where the highest bidder wins but pays the second-highest bid. Across \num{494608} auctions, winning bids averaged 2.46$\times$ (std.\ 2.79) the clearing price, with half exceeding it by at least 1.61$\times$ and ratios ranging from 1 to 306.04. Fig.~\ref{fig:timeboost-first-vs-second-price} shows that dominant controllers impose bidding limits and particularly Selini Capital and Wintermute, tend to overbid. The daily bid ratio (see Fig.~\ref{fig:timeboost-bid-ratio}) was highly volatile and exceeded 80\% in the early period, before stabilizing around 65\% from May 18, 2025 onward. \stress{Auction activity declined around March 2026 as Kairos became more dominant}. 
We attribute this shift to Kairos’ introduction of a dedicated CEX-DEX lane, which finally enables Selini and Wintermute to leverage Kairos’ infrastructure and reduce their participation in the primary auction to only when their models predict that acquiring the express lane for 60 seconds is more profitable.

\subsection{Arbitrum's Timeboost Revenue}

\begin{figure*}[tbh]
    \centering
    \begin{subfigure}{\twocolgrid}
        \centering
        \includegraphics[width=\twocolgrid]{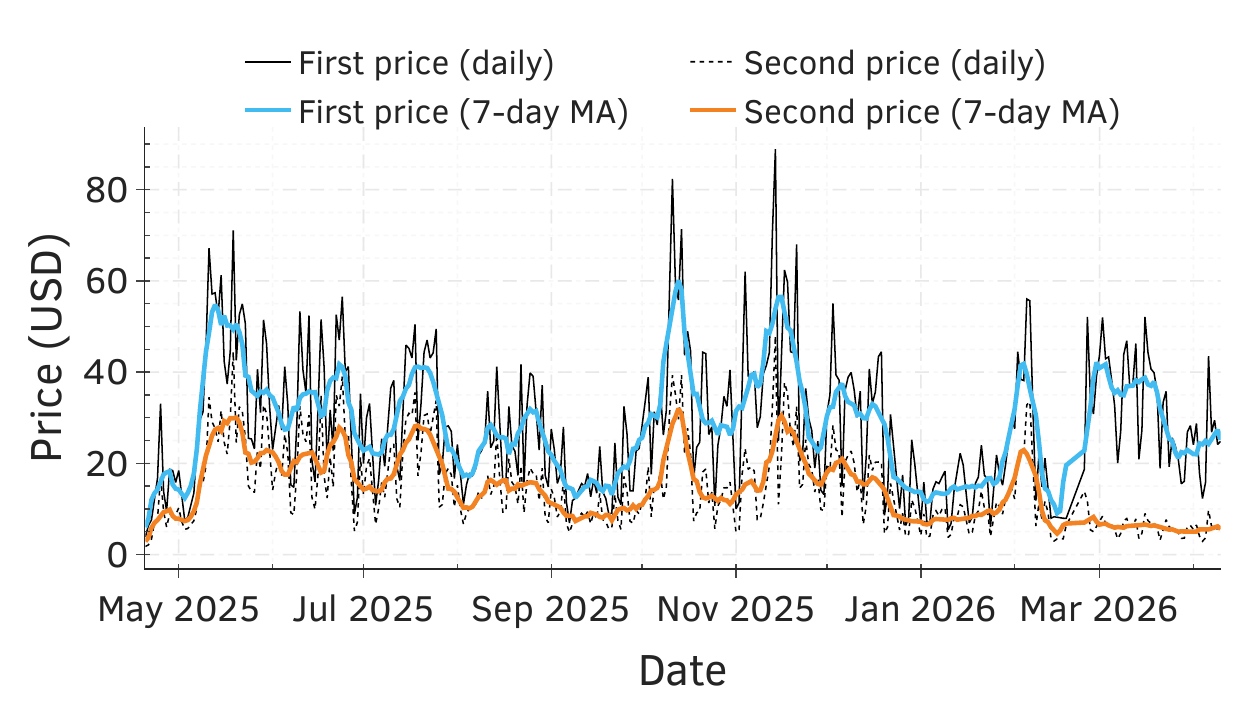}
        \caption{Auction prices over time.}
        \label{fig:timeboost-auction-prices}
    \end{subfigure}
    \begin{subfigure}{\twocolgrid}
        \centering
        \includegraphics[width=\twocolgrid]{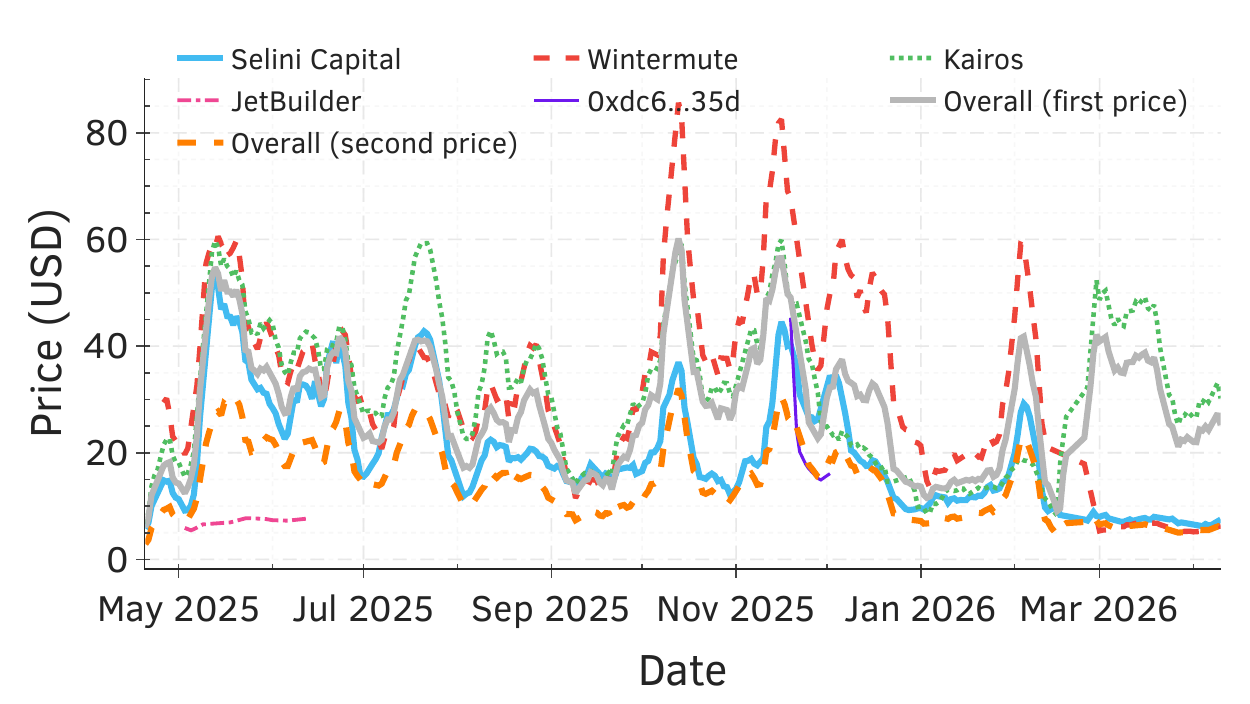}
        \caption{Moving average prices by controller.}
        \label{fig:timeboost-auction-prices-per-participant}
    \end{subfigure}\hfill
    \caption{
    Evolution of Timeboost auction prices in USD: (a) First prices decline as competition wanes and dominance consolidates; (b) Controller-level bidding reveals disciplined strategies by Selini Capital and Wintermute.
    }
    \label{fig:bids}
\end{figure*}

During our observation period, Arbitrum DAO collected \num{2442.84} ETH through Timeboost auctions, corresponding to approximately 7.1 million USD using the Binance ETH-USD rate at settlement.
Fig.~\ref{fig:timeboost-auction-prices} shows winning (first-price) and clearing (second-price) bids. As expected, winning bids consistently exceed clearing prices. 
A marked shift occurs in late February 2026, when Kairos becomes the dominant express lane controller. Clearing prices subsequently collapse. The \stress{7-day moving average falls from roughly 10--30 USD to 5.65 USD per round} and often reaches only a few dollars. Despite winning bids remaining substantially higher, DAO revenue declines sharply, indicating reduced competitive pressure.
Fig.~\ref{fig:timeboost-auction-prices-per-participant} further illustrates this shift. Earlier, Selini Capital and Wintermute competed regularly, sustaining higher auction revenues. Kairos initially placed aggressive, volatile bids but converged to the low-price regime after establishing dominance, while JetBuilder exited entirely. Together, these changes substantially reduced meaningful competition.

\subsection{Timeboost Utilization}
\label{sec:timeboost-utilization}

\begin{figure*}[tbh]
  \centering
\includegraphics[width=\textwidth]{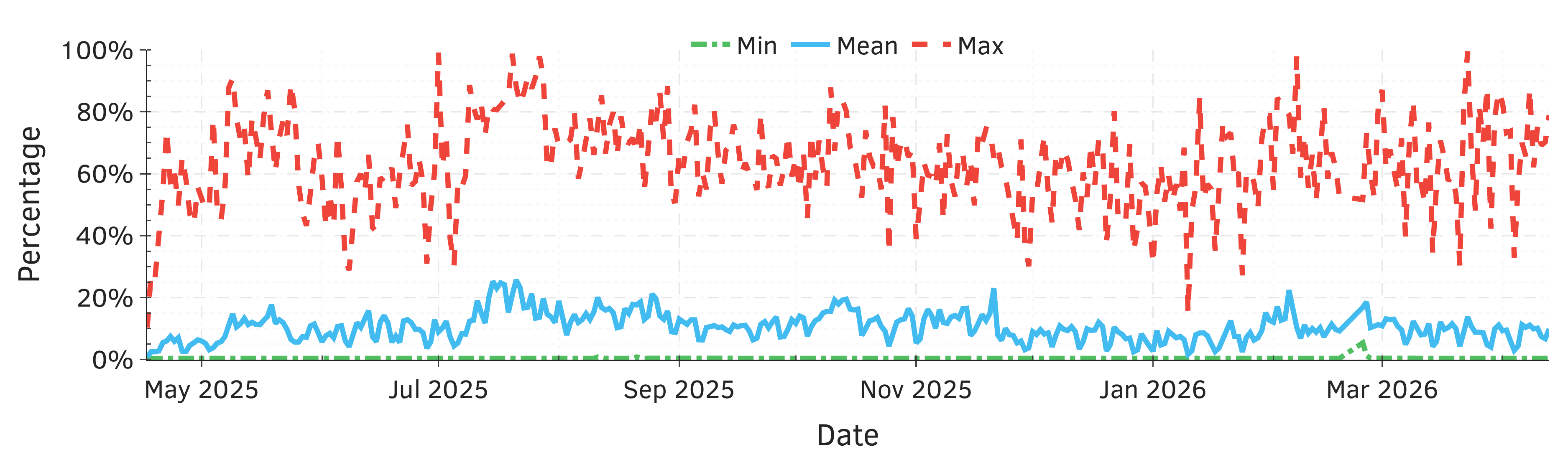}
    \caption{
    Daily minimum, mean, and maximum Timeboost utilization. Despite occasional peaks above 60\%, average utilization remains low, indicating selective rather than continuous use during one-minute control intervals.
    }
  \label{fig:timeboost-utilization-overtime-all}
\end{figure*}

Winning a Timeboost auction grants exclusive express lane access for approximately one minute, but controllers need not use it continuously. We therefore measure utilization as the fraction of blocks in each round containing at least one timeboosted transaction from the winning controller.
Across all rounds, average utilization is only 10.72\%, with the 90th and 95th percentiles at 24.27\% and 31.80\%, respectively, indicating that \stress{continuous express lane usage is uncommon}.
Utilization also varies by controller. Wintermute has the highest average at 15.68\% (median 12.50\%), followed by Kairos at 8.30\% and Selini Capital at 7.83\%. JetBuilder has the lowest at 1.41\%. Fig.~\ref{fig:timeboost-utilization-overtime-all} shows that daily mean utilization typically remains below 20\%, while daily maxima often exceed 60\%, indicating that high utilization is confined to specific rounds, and that 60-second rounds might be too large.

\subsection{Auction Bids History}
\label{sec:auction-bids-history}

\begin{figure*}[tbh]
    \centering
    \begin{subfigure}{\twocolgrid}
        \centering
        \includegraphics[width=\twocolgrid]{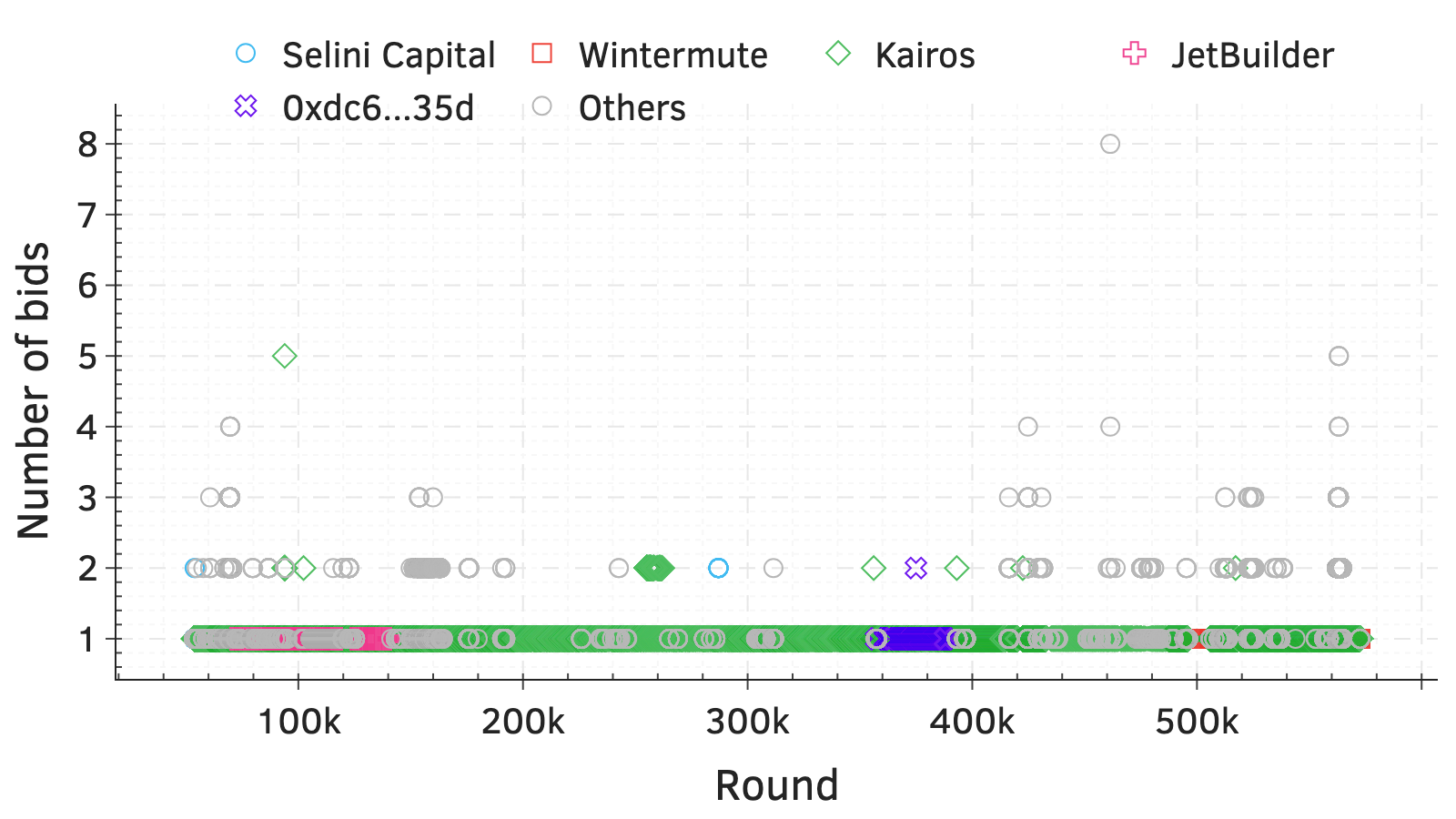}
        \caption{Number of bids per round by bidder.}
        \label{fig:timeboost-auction-bids-numbers}
    \end{subfigure}
    \begin{subfigure}{\twocolgrid}
        \centering
        \includegraphics[width=\twocolgrid]{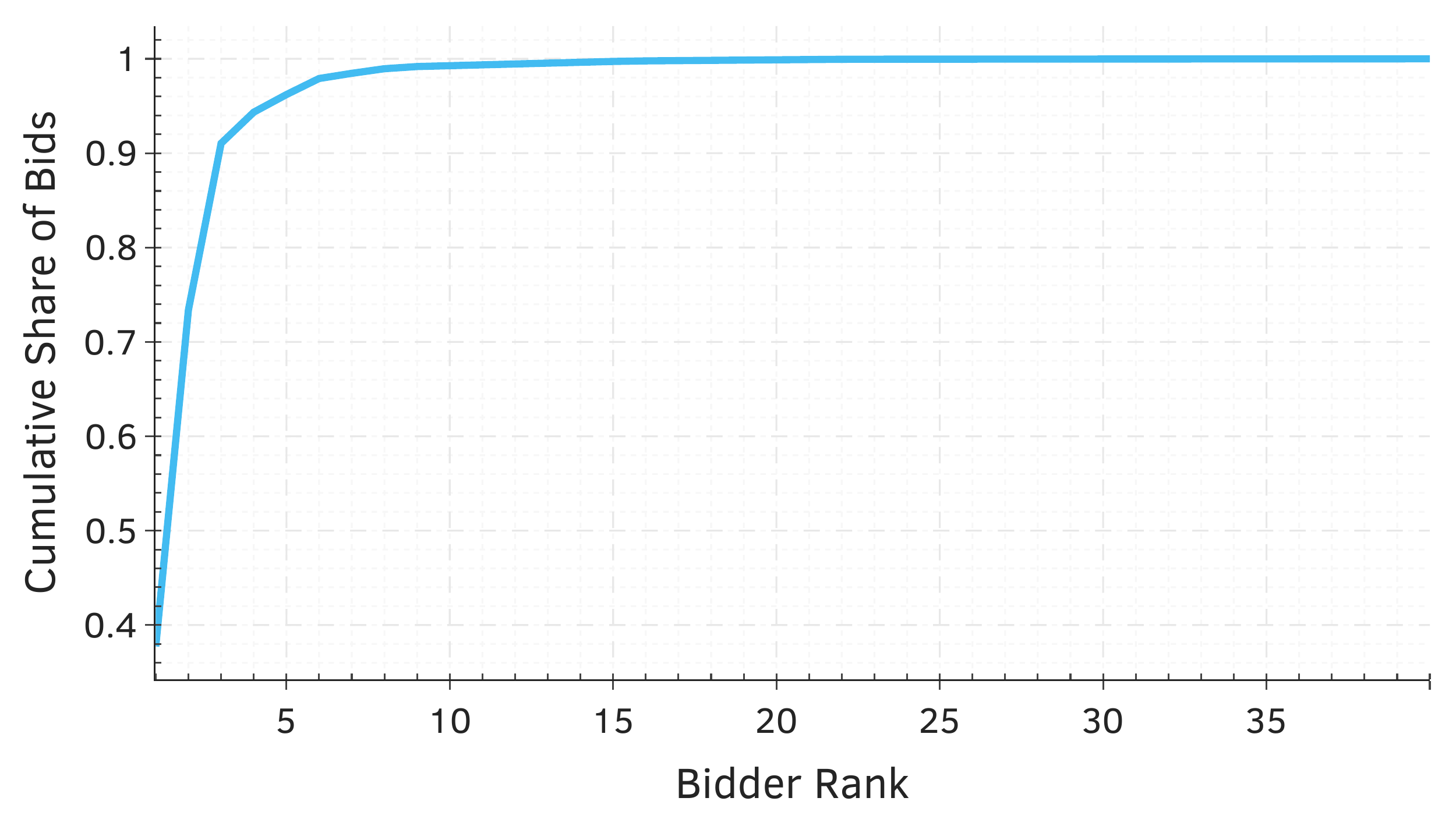}
        \caption{CDF of number of bids.}
        \label{fig:timeboost-auction-bids-concentration}
    \end{subfigure}\hfill
    \caption{Historical bidding activity in Timeboost auctions. 
    Most participants submit at most one bid per round, repeated bidding is uncommon. 
    Top three bidders account for more than 91\% of all bids.
    }
    \label{fig:timeboost-auction-bids}
\end{figure*}

We observe \num{1196302} bids from 40 unique bidder addresses, including \num{1195606} unique bid values. Of \num{494608} auction rounds, \num{492900} (99.65\%) received at least one bid, indicating near-continuous participation.
Competition within individual auctions, however, remains limited. Each round receives an average of 2.43 bids (std.=0.87), with a median of 2 and a maximum of 10. Fig.~\ref{fig:timeboost-auction-bids-numbers} shows that \stress{most bidders submit only one bid per round}, with repeated bidding being uncommon.
Bidding is also highly concentrated. Selini Capital, Wintermute, and Kairos account for 37.74\%, 35.65\%, and 17.65\% of all bids, respectively, totaling 91.04\%. Adding JetBuilder raises this to 94.35\%. Fig.~\ref{fig:timeboost-auction-bids-concentration} highlights this concentration. The top five bidders account for 96.21\% of all bids, leaving the remaining 35 bidders with less than 4\%. Thus, while participation is widespread, sustained bidding is dominated by a small group of professional actors.

\section{Timeboost's Impact on MEV Extraction} \label{sec:critical}

We examine Timeboost's impact on MEV extraction in terms of profitability, block position, controller participation, and secondary-market viability.

\begin{figure}[tbh]
    \centering
    \includegraphics[width=1.0\linewidth]{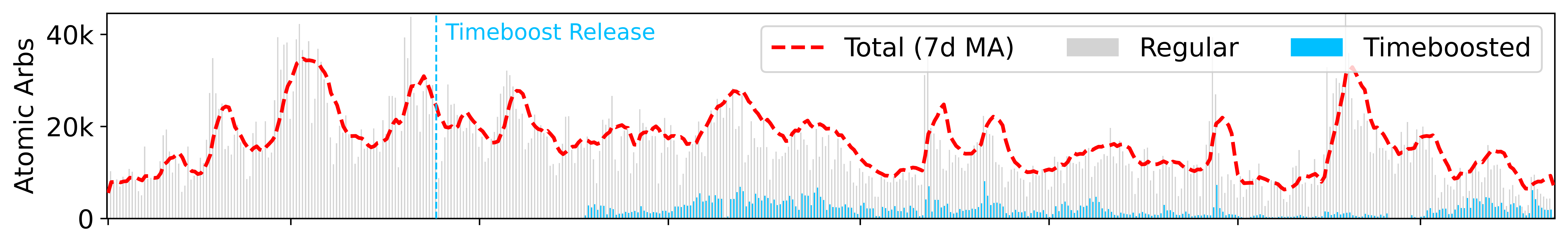}
    \includegraphics[width=1.0\linewidth]{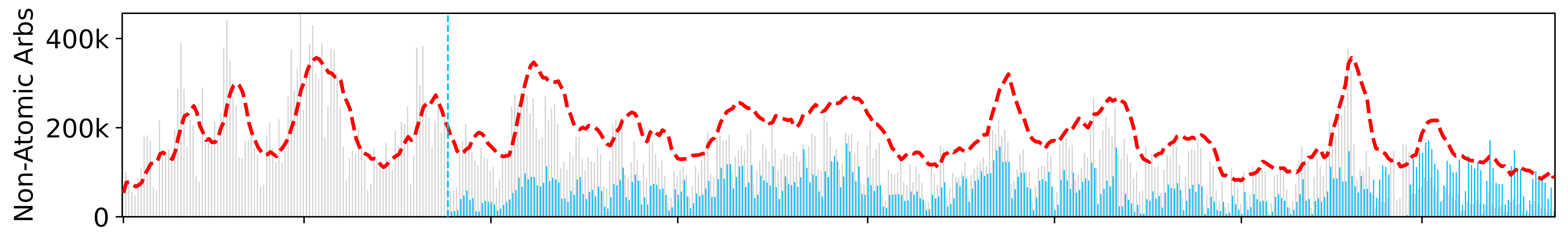}
    \includegraphics[width=1.0\linewidth]{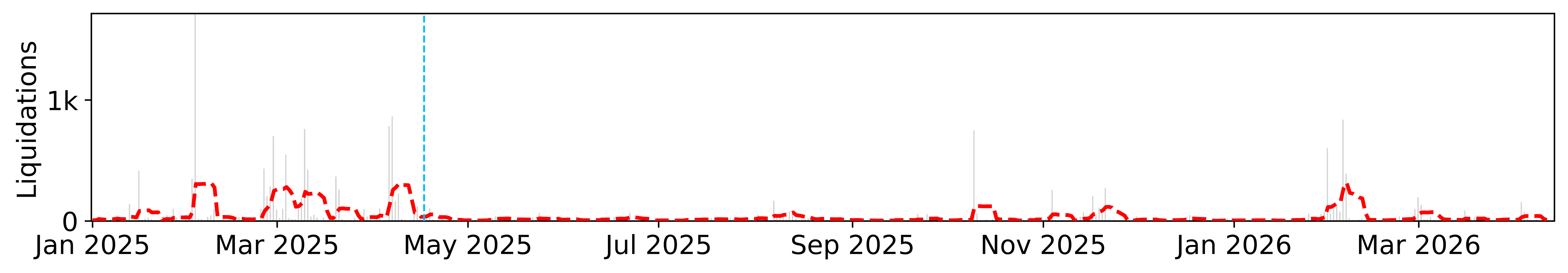}
    \caption{Number of regular and timeboosted MEV transactions over time.}
    \label{fig:arbitrages-over-time}
\end{figure}

\subsection{Timeboost Arbitrage Profits}

\figureautorefname{}~\ref{fig:arbitrages-over-time} summarizes MEV extraction before and after Timeboost's deployment, including the share of timeboosted activity. Liquidations are rare, with only \num{20299} identified, of which 99.78\% are regular and 0.22\% timeboosted. We therefore focus on atomic and non-atomic arbitrages, which account for \num{7779084} and \num{86877427} instances, respectively.
Even after a year of Timeboost, \stress{regular atomic arbitrages remain more common than timeboosted ones}, accounting for \num{5068084} (88.68\%) and \num{646953} (11.32\%) instances, respectively. In contrast, timeboosted activity is more prevalent among non-atomic arbitrages, with \num{23417412} (35.71\%) timeboosted and \num{42150241} (64.29\%) regular instances.
Interestingly, as shown in \tableautorefname{}~\ref{tab:arbitrage_profits}, \stress{timeboosted non-atomic arbitrages appear to yield higher average profits}, with a mean profit of 2.54 USD per timeboosted non-atomic arbitrage compared to 1.11 USD per regular non-atomic arbitrage. For atomic arbitrages, the opposite pattern emerges, regular atomic arbitrages yield a higher mean profit of 1.67 USD, compared to 0.91 USD for timeboosted atomic arbitrages. This difference may be explained by the higher barriers and costs associated with accessing the express lane. Purchasing express lane access directly incurs significant costs, while renting access through Kairos may also introduce additional costs that are difficult to sustain given that atomic arbitrages generally generate lower profits than non-atomic arbitrages. Atomic arbitrage opportunities are also significantly less frequent than non-atomic opportunities.

\begin{table}[t]
    \centering
    \footnotesize
    \caption{Profitability analysis of regular and timeboosted arbitrages in USD.}
    \begin{adjustbox}{max width=\columnwidth}
    \begin{tabular}{l r r r r r r}
       \toprule
       & \multicolumn{4}{c}{\textbf{Regular Arbitrage}} & & \\
       \cmidrule{2-5} & \multicolumn{2}{c}{\textbf{Pre-Timeboost}} & \multicolumn{2}{c}{\textbf{Post-Timeboost}} & \multicolumn{2}{c}{\textbf{Timeboosted Arbitrage}} \\ 
       \cmidrule{2-7}
       & \multicolumn{1}{c}{\textbf{Non-Atomic}} & \multicolumn{1}{c}{\textbf{Atomic}} & \multicolumn{1}{c}{\textbf{Non-Atomic}} & \multicolumn{1}{c}{\textbf{Atomic}} &
       \multicolumn{1}{c}{\textbf{Non-Atomic}} & \multicolumn{1}{c}{\textbf{Atomic}} \\
       \midrule
       Total    & \num{15212835.09} & \num{1780245.10} & \num{22782750.22} & \num{6970910.30} & \num{31214567.08} & \num{533656.56} \\
       Max      & \num{4652.60} & \num{59684.66} & \num{22887.82} & \num{49988.01} & \num{19785.10} & \num{71756.57} \\
       Mean     & \num{1.31} & \num{0.91} & \num{1.11} & \num{1.67} & \num{2.54} & 0.91 \\
       Median   & \num{0.16} & \num{0.02} & \num{0.11}  &  \num{0.03} & \num{0.43} & 0.02 \\
       \bottomrule
    \end{tabular}
    \end{adjustbox}
    \label{tab:arbitrage_profits}
\end{table}

\begin{table}[b]
    \centering
    \caption{Timeboosted arbitrages, arbitrageurs, and bots per controller.}
    \begin{adjustbox}{max width=\columnwidth}
    \begin{tabular}{l r r r r r r}
        \toprule
         & \multicolumn{2}{c}{\textbf{Arbitrages}} & \multicolumn{2}{c}{\textbf{Arbitrageurs}} & \multicolumn{2}{c}{\textbf{Arbitrage Bots}} \\
         \cline{2-7} \noalign{\vskip 3pt}
        \textbf{Controller} & \textbf{Non-Atomic} & \textbf{Atomic} & \textbf{Non-Atomic} & \textbf{Atomic} & \textbf{Non-Atomic} & \textbf{Atomic} \\
        \midrule
        Wintermute & \num{13360379} & 0 & 115 & 0 & 4 & 0 \\
        Selini Capital       & \num{5711966} & \num{519334} & 410 & 85 & 4 & 14 \\
        Kairos               & \num{4344871}  & \num{123989} & 436 & 146 & 4 & 47 \\
        JetBuilder           & 164 & 649  & 10 &  1 & 1 & 1 \\
        \bottomrule
    \end{tabular}
    \end{adjustbox}
    \label{tab:arbitrages}
\end{table}

\subsection{Express Lane Controllers}
Table~\ref{tab:arbitrages} summarizes MEV activity among the top controllers. Wintermute is the only controller that does not engage in atomic arbitrage, instead leading non-atomic arbitrage with over 13 million instances across four bots and 115 sender addresses. Selini Capital performs over 5 million non-atomic arbitrages and leads timeboosted atomic arbitrage with over 500 thousand instances.
Kairos engages in both types of arbitrage too, but its activity changed substantially after February 2026, when Wintermute and Selini Capital began using Kairos. This explains its 436 distinct sender addresses, which largely comprise accounts associated with these two entities, as well as its more than 4 million non-atomic arbitrages. Unlike Selini Capital, Kairos also has a large number of distinct bots, likely because atomic arbitrage is performed by multiple Kairos users, whereas non-atomic arbitrage is primarily driven by Wintermute and Selini Capital. Table~\ref{tab:controller_arbitrage_profits} shows that \stress{atomic arbitrage generates substantially lower total profits than non-atomic arbitrage}. Comparing these profits with the fees controllers pay to Arbitrum to purchase the express lane, we find that atomic arbitrage alone would not cover the cost of express lane access, whereas non-atomic arbitrage could. This suggests that there is currently a limited economic viability for atomic arbitrage through Timeboost on Arbitrum.

\begin{table}[t]
    \centering
    \caption{Total fees paid per controller compared to arbitrageur profit in ETH.}
    \begin{adjustbox}{max width=\columnwidth}
    \begin{tabular}{l r r r r r r r}
        \toprule
        &  & \multicolumn{2}{c}{\textbf{Total Profit}} & \multicolumn{2}{c}{\textbf{Max Profit}} 
        & \multicolumn{2}{c}{\textbf{Median Profit}} 
        \\
        \cmidrule{3-8}
        \textbf{Controller} & \textbf{Total Fees} & \textbf{Non-A.} & \textbf{Atomic} & \textbf{Non-A.} & \textbf{Atomic} & \textbf{Non-A.} & \textbf{Atomic} \\
        \midrule
        Wintermute & \num{1079.26} & \num{2267.16} & 0 & 3.82 & 0 & -0.0000004 & 0  \\
        Selini Capital & \num{1008.95} & \num{2531.51} &  116.18 & 5.73 & 16.65 & 0.00003 & 0.000004 \\
        Kairos & \num{301.28} & 647.87 & 17.72 & 2.08 & 0.42 & 0.0000004 & 0.000006 \\
        JetBuilder & \num{2.89} & 0.04 & 0.94 & 0.003 & 0.14 & 0.00005 & 0.0001 \\
        \bottomrule
    \end{tabular}
    \end{adjustbox}
    \label{tab:controller_arbitrage_profits}
\end{table}

\subsection{Timeboost Arbitrage Positions}

\begin{figure*}[tbh]
    \centering
    \includegraphics[width=1.0\linewidth]{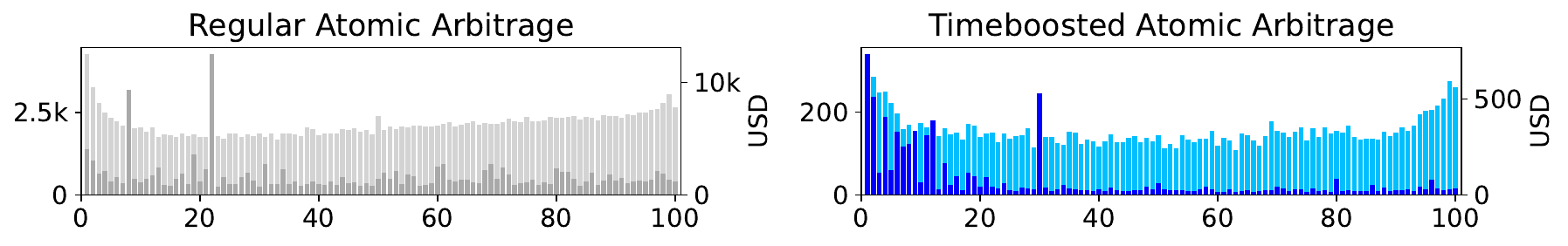}
    \includegraphics[width=1.0\linewidth]{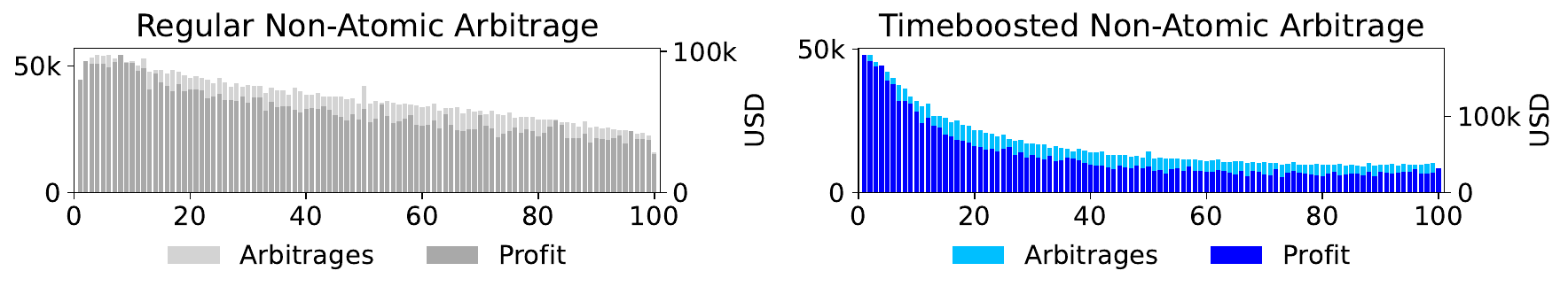}
    \caption{Distribution of regular and timeboosted arbitrage transaction  positions.}
    \label{fig:arbitrage-transaction-positions}
\end{figure*}

Due to Arbitrum's short 250 ms block time, blocks are relatively small, with median sizes of 13 transactions for blocks containing atomic arbitrages and 20 for non-atomic arbitrages. To better understand Timeboost's priority access, we analyze the positions of regular and timeboosted transactions in blocks with at least 100 transactions, focusing on periods of high blockspace demand.
\figureautorefname{}~\ref{fig:arbitrage-transaction-positions} shows the positions of regular atomic and non-atomic arbitrages (left) and timeboosted arbitrages (right). Darker bars indicate the aggregate USD profit extracted at each block position.
\stress{Timeboost appears to provide stronger top-of-block priority for non-atomic arbitrage}, whereas atomic arbitrages frequently occur near the end of the block regardless of whether they are timeboosted. This may reflect the additional latency introduced by Kairos' sub-auction mechanism and the lower baseline latency of its users. In contrast, large participants such as Selini Capital and Wintermute likely already operate with highly optimized latency, helping explain the strong top-of-block positioning of non-atomic arbitrages.
For non-atomic arbitrage, positions near the top of the block also tend to yield the highest profits, consistent with their dependence on off-chain prices and the latest on-chain state. Atomic arbitrage exhibits a somewhat different pattern, although early positions are profitable, high profits can also occur slightly later in the block, potentially because atomic arbitrage can exploit price changes created by preceding transactions, such as by non-atomic arbitrages.
These differences suggest a potential design with two dedicated express lanes instead of one: one lane prioritizing non-atomic arbitrage and another ordering atomic arbitrage immediately afterward.

\subsection{Economic Viability of Secondary Markets} 
\begin{figure}[tbh]
    \centering
    \includegraphics[width=1.0\linewidth]{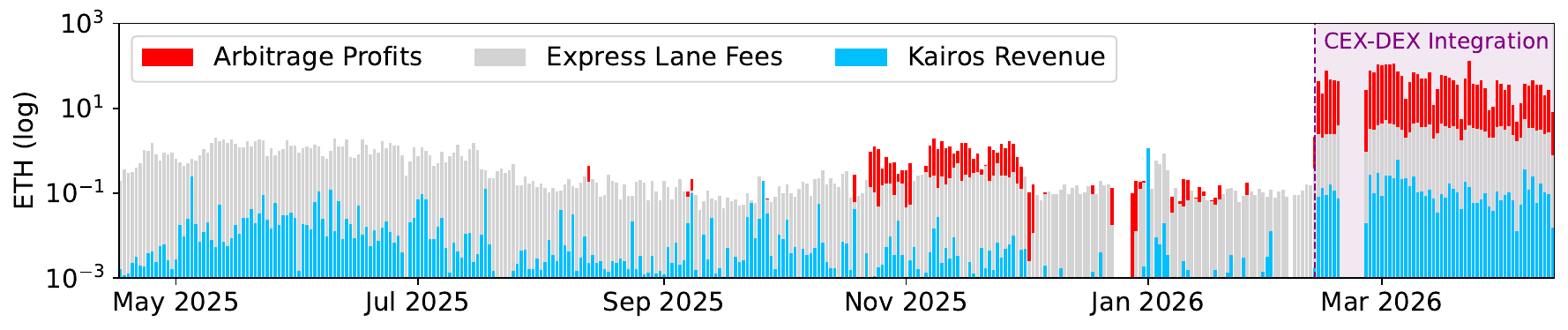}
    \caption{Arbitrageur profits, express lane fees, and Kairos revenue over time.}
    \label{fig:kairos-profitability}
\end{figure}

Kairos received 349,076 payments from 131 contracts, totaling 11.96 ETH, with a mean payment of 0.00003 ETH. Notably, 78,906 payments (22.6\%) were only 1 wei (Kairos' minimum bid~\cite{Kairos-Docs}), while 6,045 (1.7\%) paid no fee even. In contrast, Kairos paid 301.28 ETH in fees for express lane access, with a mean of 0.004 ETH per round.
Fig.~\ref{fig:kairos-profitability} shows that before the CEX-DEX integration, Kairos generally paid substantially more in fees than it collected, while arbitrageurs often earned more than both Kairos’ revenue and, at times, its express lane fees. Overall, \stress{Kairos was not profitable before the CEX-DEX integration}, raising once again the question whether controllers that do not engage in non-atomic arbitrage can sustain the cost of express lane access. After the CEX-DEX integration, both revenue and fees increased. Because CEX-DEX payments to Kairos occur off-chain, we cannot directly assess its profitability. However, aggregate arbitrage (i.e., non-atomic and atomic) profits appear sufficient to cover Kairos' express lane fees, assuming arbitrageurs would pay Kairos a fair share.

\section{Discussion} \label{sec:discussion}

Our results reveal a fundamental tension between Timeboost's design goals and observed outcomes. 
We discuss directions for future improvement.

\medskip
\noindent
\textbf{Controlling Centralization.}
Although Timeboost rotates express-lane rights each round, a few entities, primarily Selini Capital, Wintermute, and later Kairos, dominate the auctions. This concentration may reduce participation diversity, enable collusion, and undermine long-term sustainability by pricing smaller actors out of the market.
Several mechanisms could broaden participation. Participation caps could limit repeated wins within a sliding window, although they are vulnerable to Sybil attacks. Whitelisting, identity-bound participation, staking, or reputation mechanisms could mitigate the risk of Sybil attacks at the cost of greater permissioning. Auction-level changes could also reduce concentration, such as randomized allocation among the top-$k$ bidders~\cite{Celis-WWW11}, more frequent, dynamically adjusted reserve prices based on participation and concentration, or allowing multiple controllers to share express-lane capacity within the same round. More broadly, Timeboost could optimize for both revenue and participation diversity, as our results show that short-term revenue gains diminish as centralization reduces competition.

\medskip
\noindent
\textbf{Drawing on Secondary Markets.}
Secondary markets can improve efficiency by reallocating rights to participants who value them most, but may also encourage speculation and concentrate control among well-capitalized intermediaries. We observe Wintermute and Selini Capital shifting their MEV extraction from direct Timeboost participation to Kairos following the introduction of its dedicated CEX-DEX lanes. Kairos's growth reflects demand for per-transaction priority and its redesigned pipeline, which bypasses simulation to reduce latency but can increase failed transactions due to insufficient dependency checks. Timeboosted transactions frequently revert, particularly those routed through Kairos. We find that flaws in Kairos' transaction bundling potentially allows costless spam in the form of purposely reverted transactions at the top of blocks, enabling block-stuffing (DoS) attacks.
Rather than relying on opaque secondary markets, Timeboost could offer similar dynamics through shorter rounds via standardized APIs offering better execution guarantees in the form of atomic bundles. Shorter rounds would lower participation costs, potentially attract more bidders, and increase participation diversity without necessarily reducing DAO revenue.

\section{Related Work} \label{sec:related_work}

\parab{MEV and Fair Transaction Ordering}
Daian et al.~\cite{daian2020flash} were the first to formalize \gls{MEV}, showing how privileged actors extract value by reordering, inserting, or censoring transactions and highlighting its implications for market fairness and consensus security. Existing mitigations primarily target mempool privacy or order fairness. Threshold encryption~\cite{Choudhuri2024ThresholdEncryption} hides transaction contents until ordering decisions are finalized, while Wendy~\cite{Kursawe2020Wendy}, Aequitas~\cite{Kelkar-Aequitas}, Themis~\cite{Kelkar2023Themis}, and Helix~\cite{Yakira2021Helix} provide cryptographic or consensus-level ordering guarantees. However, these approaches often introduce complexity, latency, and communication overhead, or require multi-sequencer architectures.
Rollups exhibit similar but distinct \gls{MEV} dynamics due to their private mempools~\cite{Torres@CCS24} and single-sequencer systems. Gogol et al.~\cite{gogol2024crossrollupmevnonatomicarbitrage,Gogol@MARBLE24} show that fragmented liquidity enables cross-rollup arbitrage, highlighting how ordering mechanisms shape value distribution. Yet empirical studies of deployed ordering mechanisms in rollups remain scarce.

\parab{Auction-Based Ordering and Timeboost}
An alternative allocates ordering rights through auctions. On Ethereum, \gls{PBS}~\cite{Heimbach@IMC23} lets builders compete for block-construction rights, directing \gls{MEV} revenue toward validators and other protocol participants. Similar mechanisms have been proposed for rollups. Arbitrum's Timeboost~\cite{arbitrum2025timeboost,Mamageishvili2023Timeboost} extends \gls{FCFS} ordering with a sealed-bid second-price auction for temporary ``express lane'' access, aiming to reduce latency races and internalize \gls{MEV} for Arbitrum's DAO.
Recent work examines Timeboost's economics. Fritsch et al.~\cite{Fritsch2024Capture} show that auction-based ordering can internalize \gls{MEV} otherwise captured by private actors, while Mamageishvili et al.~\cite{mamageishvili2025timeboostaheadoftimeauctionswork} find that users have difficulties in pricing short-term ordering opportunities. Öz et al.~\cite{oz2026jit} show that just-in-time markets, such as Kairos~\cite{Kairos}, reduce primary-auction competition and revenue captured by Arbitrum. However, Timeboost's effects on centralization, MEV extraction, and transaction efficiency remain largely unexplored. We address this gap with the first large-scale study of Timeboost, showing that demand for priority access is dominated by CEX--DEX arbitrage and that secondary markets can remain viable only by supporting such activity. However, this may further increase centralization, as low participation can enable secondary-market intermediaries to bypass the primary auction and reduce DAO revenue over time. We also uncover inefficiencies in Kairos' transaction-ordering guarantees that may increase transaction reverts and, consequently, spam.

\section{Conclusion} \label{sec:conclusion}

This paper presents the first large-scale empirical study of Timeboost, analyzing \num{48525587} timeboosted transactions and \num{494608} auctions from its launch on April 17, 2025, to April 13, 2026. We find that three entities, Selini Capital, Wintermute, and Kairos, won over 99.74\% of auctions and account for most timeboosted transactions. Despite earlier inclusion, nearly one in three express lane transactions revert. Selini Capital and Wintermute primarily use Timeboost for CEX-DEX arbitrage, limiting the value of priority access for other MEV opportunities. Secondary markets also proved difficult to sustain: JetBuilder exited, while Kairos introduced dedicated CEX-DEX lanes to accommodate these dominant participants. Consequently, Selini Capital and Wintermute increasingly access the express lane indirectly through Kairos rather than participating in primary auctions, reducing auction revenue for the Arbitrum DAO.
Overall, Timeboost primarily serves latency-sensitive CEX-DEX arbitrageurs while failing to reduce spam or broaden participation. Its current design instead reinforces concentration and narrows its practical use cases, highlighting a gap between the intended benefits of auction-based ordering and its outcomes in practice.

\section*{Acknowledgments}\label{sec:ack}

The authors gratefully acknowledge the support of the Dune team, whose provision of platform access was instrumental to the research presented in this paper.
This work is also supported by the European Union’s Horizon 2020 research and innovation programme under grant agreement No 952226, project BIG (Enhancing the research and innovation potential of Tecnico through Blockchain technologies and design Innovation for social Good) as well as by national funds through FCT, Fundação para a
Ciência e a Tecnologia, under projects UIDB/500
21/2020 (DOI:10.54499/UIDB/50021/2020) and UIDP/50021/2020 (DOI:10.544
99/UIDP/50021/2020).

%------------------------------------------------------------------------------

\bibliographystyle{splncs04}
\bibliography{references}

\appendix

\section{Data Collection}
\label{sec:data-collection}

\textbf{Timeboost Auctions and Controllers.}
We extract all rounds of the Timeboost sealed-bid second-price auctions from the on-chain Timeboost contract.\footnote{Arbitrum's Timeboost contract is deployed at contract address: \href{https://arbiscan.io/address/0x5fcb496a31b7ae91e7c9078ec662bd7a55cd3079}{0x5fcb$\cdots$3079}.} For each round we record the winning bidder, the submitted first-price bid, the clearing (second-price) payment, and the designated express lane controller address along with the time the controller is allowed to submit their transactions via the express lane. In total, our dataset contains \num{494608} auctions with 32 unique winning controllers. These bids where made by \num{40} unique addresses.

\medskip
\noindent
\textbf{Historical Bid Data.}
To complement the on-chain auction outcomes, we also collect Arbitrum's historical bid trace dataset, which publicly exposes the sequence of submitted bids for each auction round. The dataset was retrieved from the official Timeboost bid history archive provided by the Arbitrum Foundation~\cite{Arbitrum-Bids}. Concretely, the archive contains compressed per-round bidding logs hosted in a public cloud storage bucket, where each entry records information such as the auction round identifier, bidder address, bid value, timestamp, and auction outcome. Overall, our bid history dataset contains \num{1196302} bid submissions from 40 unique bidder addresses, including \num{1195606} unique bids.

\medskip
\noindent
\textbf{Timeboosted Transactions.}
To evaluate the transactions submitted through the express lane by auction winners, we filtered all timeboosted transactions during our study period. In total, \num{48525587} transactions were sent through Timeboost's express lane, accounting for \num{4.52}\% of all transactions included in Arbitrum's blocks from the moment timeboost was launched. These transactions originated from 1,240 unique accounts and were submitted to the express lane by \num{32} different auction winners. This highlights secondary markets such as \emph{Kairos}, which submit users' transactions to the express lane on their behalf. These transactions target 165 unique `to' addresses, 153 of which are contracts.

\medskip
\noindent
\textbf{\gls{MEV} Extraction.}
To analyze \gls{MEV} activity during our study period, we build on the MEV detection framework proposed by Torres et al.~\cite{Torres@CCS24}. We extended their scripts by adding support for detecting Uniswap V4 and PancakeSwap trades. 
For non-atomic arbitrage detection, we combine the approaches of Heimbach et al.~\cite{Heimbach-Non-Atomic-Arbitrage@SP24} and Öz et al.~\cite{oz2026jit}. To limit false positives, we restrict detection to single-swap trades involving WETH, WBTC, USDC, or USDT, which are highly liquid and actively traded across CEXes and DEXes. However, as opposed to \cite{oz2026jit}, we do not leverage a set of hard-coded arbitrageur addresses, thereby catching more arbitrageurs. We further require that transactions are not previously identified as atomic arbitrages or liquidations, that the swap is the first swap in the respective direction and DEX pool within the block, and that the ``to'' address is a contract without verified source code on Sourcify~\cite{sourcify}. Finally, we exclude detected transactions associated to addresses with fewer than 100,000 detected non-atomic arbitrages. We compute profits and costs using Binance markout prices denominated in USDT over a 10-second window, selecting the window with the highest resulting profit to provide an estimate.
Using this approach, we identify \num{7779084} cyclic atomic arbitrage transactions, \num{86877427} non-atomic arbitrage transactions, and \num{20299} liquidation transactions on Arbitrum. Of the cyclic arbitrages, \num{7132131} (91.7\%) are regular and \num{646953} (8.3\%) are timeboosted. For non-atomic arbitrages, \num{63460015 } (73.05\%) are regular and \num{23417412} (26.95\%) are timeboosted. Finally, \num{20254} (99.8\%) liquidations are regular and \num{45} (0.2\%) are timeboosted.

\medskip
\noindent
\textbf{Kairos Revenue.}
We collected via Dune's traces table all payments made to Kairos,\footnote{Kairos's payment contract is deployed at address: \href{https://arbiscan.io/address/0x60e6a31591392f926e627ed871e670c3e81f1ab8}{0x60e$\cdots$ab8}.} along with metadata such as payee addresses and payment timestamps. In total, we identified \num{349076} user payments associated with express lane usage.

\medskip
\noindent
\textbf{Exchange Rates.}
To compute the US dollar price at the time each transaction or auction bid was issued, we use Binance's USDT exchange price data~\cite{Binance-API} sampled at 1-second intervals.

\end{document}